\documentclass[12pt]{article}
\usepackage{epsfig}
\usepackage{amssymb}
\usepackage{amsmath}
\usepackage{amsfonts}
\usepackage{graphicx}
\usepackage{mathrsfs}
\usepackage[dvips]{color}
\usepackage{multirow}

\newcommand{\fc}{\mathfrak{c}}

\newcommand{\fe}{\mathfrak{e}}

\newcommand{\fx}{\mathfrak{x}}

\newcommand{\fz}{\mathfrak{z}}

\newcommand{\fI}{\mathfrak{I}}

\newcommand{\fK}{\mathfrak{K}}

\newcommand{\bL}{\mathbf{L}}

\newcommand{\bS}{\mathbf{S}}
\newcommand{\bT}{\mathbf{T}}

\newcommand{\cD}{\mathcal{D}}

\newcommand{\cI}{\mathcal{I}}

\newcommand{\cK}{\mathcal{K}}
\newcommand{\cL}{\mathcal{L}}
\newcommand{\cM}{\mathcal{M}}

\newcommand{\cP}{\mathcal{P}}

\newcommand{\cS}{\mathcal{S}}

\newcommand{\cZ}{\mathcal{Z}}
\newcommand{\be}{\begin{equation}}
\newcommand{\ee}{\end{equation}}
\newcommand{\bea}{\begin{eqnarray}}
\newcommand{\eea}{\end{eqnarray}}
\newcommand{\nn}{\nonumber}

\newcommand{\ed}{\end{document}}

\newcommand{\bi}{\begin{itemize}}
\newcommand{\ei}{\end{itemize}}

\newcommand{\bce}{\begin{center}}
\newcommand{\ece}{\end{center}}

\newcommand{\sL}{\mathscr{L}}

\newcommand{\sX}{\mathscr{X}}
\newcommand{\sY}{\mathscr{Y}}

\newcommand{\p}{\partial}

\newcommand{\md}{{(\cM_D,g)}}
\newcommand{\mink}{(\cM_D,\eta)}

\begin{document}

\title{Quantum of the bare cosmological constant}

\author{Farhang Loran\thanks{E-mail address: loran@iut.ac.ir} \\[6pt] Department of Physics, Isfahan University of Technology, \\ Isfahan
84156-83111, Iran\\[6pt] }

\date{ } \maketitle

\begin{abstract}
We show that there exist  scalar field theories  with  plausible  one-particle states in general $D$ dimensional nonstationary curved spacetimes whose propagating modes  are localized on $d\le D$ dimensional hypersurfaces, and the corresponding stress tensor resembles  the bare cosmological constant $\lambda_B$ in  the $D$ dimensional bulk. We show that nontrivial  $d=1$ dimensional solutions  correspond to $\lambda_B< 0$. Considering free scalar theories we find that for $d=2$  the symmetry of the parameter space of classical solutions corresponding to $\lambda_B\neq 0$ is $O(1,1)$ which enhances to $\mathbb{Z}_2\times{\rm Diff}(\mathbb{R}^1)$ at $\lambda_B=0$. For $d>2$ we obtain $O(d-1,1)$,  $O(d-1)\times {\rm Diff}(\mathbb{R}^1)$  and $O(d-1,1)\times O(d-2)\times {\rm Diff}(\mathbb{R}^1)$ corresponding to, respectively, $\lambda_B<0$,  $\lambda_B=0$ and $\lambda_B>0$. \\

%\noindent PACS number: 04.62.+v, 11.30.Rd, 11.25.Hf, 11.25.Pm \vspace{2mm}
% Keywords: QFT in curved spacetime; one-particle states; localization to hypersurfaces; bare cosmological constant
%\medskip
 \end{abstract}
\tableofcontents
%\maketitle
\section{Introduction}\label{section-Int}
In this paper we propose  novel  scalar field theories, the scalar creepers, in general nonstationary curved spacetime whose stress tensor resemble a bare cosmological constant and the corresponding  one-particle states are localized on lower dimensional hypersurfaces.

 The particle interpretation of states of \emph{ordinary} quantum field theory (QFT) in Minkowski spacetime can be extended to QFT in stationary curved spacetime.  However, there is a lore that a physically meaningful notion of particles do not exist for QFT in a general nonstationary curved spacetime  except in an approximate or asymptotic sense. So, some authors  argue that the particle interpretation of states should not be considered as an essential feature of QFT \cite{Wald-Book}. The scalar creepers do not describe ordinary matter field but they have well-defined one particle states in general.

 Another question in QFT is the existence of matter fields whose propagating modes are localized on $d<D$ dimensional hypersurfaces of the $D$-dimensional spacetime \cite{Csaki,Bajc:1999mh}. We show that the propagator of the one-particle states of the scalar creepers are delta-function localized  on  $d$ dimensional  hypersurfaces. In general $d\le d_*$, where $d_*$ is the number of  linearly independent nowhere-zero vector fields in the spacetime which can be computed in the homotopy theory \cite{Adams}.

 The scalar creepers  are not ordinary matter fields. Their stress tensor resembles a bare cosmological constant, i.e.,  they all act like perfect fluid with equation of state $w=-1$.  An unsettled issue  in QFT in curved spacetime is the cosmological constant problem \cite{no-go}.

 Considering a general four dimensional nonstationary curved spacetime $(\cM_4,g)$  where $\cM_4$ is a smooth four-dimensional manifold and $g$ is a Lorentzian metric on $\cM_4$ with signature $(-,+,+,+)$,  local Lorentz symmetry implies that the expectation value of the quantum vacuum stress-energy tensor of ordinary fields $\left\langle T_{\mu\nu}\right\rangle=-\left\langle \rho\right\rangle g_{\mu\nu}$  where $\rho\sim\Lambda^4$ is the vacuum energy density and $\Lambda$ is the high energy cutoff of the ordinary QFT. So $\left\langle T_{\mu\nu}\right\rangle$ adds  $8\pi G\left\langle \rho\right\rangle\sim M_{\rm Pl}^{-2}\Lambda^4$ to the effective cosmological constant whose observed value is $\lambda_{\rm eff}\sim 10^{-122}M_{\rm Pl}^2$. Since $\lambda_{\rm eff}=\lambda_B+8\pi G\left\langle \rho\right\rangle$ this requires an incredible fine-tuning of $\lambda_B$.  Presumably the  cosmological constant problem can be  resolved if fluctuations of $\rho$ are taken into account and  $\lambda_B$  has   taken a large negative value $-{\lambda_B}\gg \Lambda^2$ \cite{Wang,Unruh}.
 We show the  stress tensor  of $d=1$ dimensional creepers    resembles $\lambda_B<0$, and in the simplest $d\ge2$ dimensional  models  we find that the symmetry of the parameter space of classical solutions corresponding to $\lambda_B< 0$ is $O(d-1,1)$ which enhances to $O(d-1)\times{\rm Diff}(\mathbb{R}^1)$ at $\lambda_B=0$, indicating a phase transition.

 In summary, the scalar creepers have the following properties:
 \begin{enumerate}
\item Similarly to ordinary scalars, they are natural extensions of scalars in Minkowski spacetime to curved spacetime. Their actions are diffeomorphism invariant.

\item They have a reasonable notion of one-particle states in nonstationary curved spacetimes,  localized to $d\le D$ dimensional hypersurfaces without using warp factors or potential wells, hence the moniker.

\item Their stress tensor resembles a bare cosmological constant, i.e.,  they all act like perfect fluid with equation of state $w=-1$. So they do not describe ordinary matter field.
\end{enumerate}

 In the next section we argue that  scalar creepers are  plausible entities  in time-orientable spacetimes. In section \ref{section-Minkowski} we  define scalar creepers in Minkowski spacetime, and examine their existence in general nonstationary curved spacetimes  in section \ref{section-curved}. We discuss their application to cosmology in section \ref{section-cosmology} and recapitulate the main results in section \ref{section-conclusion}.

\section{Outline}\label{section-Prologue}
In  a $D$ dimensional spacetime $\md$ whose metric is $g$ with signature $(-, +,  \cdots, +)$ a scalar field theory is given by the action
    \be
    \bS:=\int d^D y\,\fe\, \bL,
    \ee
where $\fe:=\sqrt{\left|\det g\right|}$. The kinetic term in the Lagrangian $\bL$, which we denote by $\mathbf{K}$,  is a combination of the scalar field $\varphi(y)$ and its derivatives $\p_\mu\varphi(y)$ in which  $\p_\mu:=\frac{\p}{\p y^\mu}$. This combination should be diffeomorphism invariant. For ordinary matter field  (the reciprocal of) the spacetime metric is used to prepare a suitable kinetic term:
    \be
    \mathbf{K}:= -\frac{1}{2}g^{\mu\nu}\p_\mu\varphi\p_\nu\varphi.
    \ee
If $\md$ is time-orientable then in addition to $g$ there also exists a smooth nonvanishing timelike vector field $v_0$  \cite{Wald-relativity} which can be used to write a diffeomorphism invariant ``kinetic term''
    \be
    \cK(\phi;v_0):= \frac{1}{2}v_0[\phi]^2=\frac{1}{2}{v_0}^\mu{v_0}^\nu\p_\mu\phi\p_\nu\phi,
    \ee
where $v_0[\phi]:={v_0}^\mu\p_\mu\phi$.
We call $\phi$  ``one-dimensional creeper.'' For the moment we do not consider self-interactions and possible coupling to ordinary matter field and suppose that the action is given by $ \cS:=\int d^D y\,\fe\, \cK(\phi;v_0)$. This model has a remarkable feature:  since $\cK(\phi;v_0)$  is independent of the metric,  the corresponding stress  tensor given by
    \be
    T_{\mu\nu}:=-\frac{2}{\fe}\frac{\delta\cS}{\delta g^{\mu\nu}}=\cK(\phi;v_0)\,g_{\mu\nu},
    \ee
resembles a perfect fluid with equation of state $w=-1$ similarly to the cosmological constant.  In fact, a consistent coupling to gravity requires local conservation of the stress tensor. Denoting the covariant derivative by $\nabla$ corresponding to the Levi-Civita connection we note that
    \be
     \nabla_\mu {T^\mu}_\nu=\p_\nu\cK(\phi;v_0).
    \ee
Thus $\cK(\phi;v_0)$ needs to be constant on-shell, whose value can interpreted as a ``negative'' bare cosmological constant. Thus it is plausible to investigate the creepers further.

The space of classical solutions $\phi_{\rm cl}$ such that $\cK(\phi_{\rm cl};v_0)$ is constant can be found by solving the classical equation of motion indicated by  the principle of least action ${\delta \cS}/{\delta\phi}=0$. The solutions to this field equation are highly nonunique because $\cK(\phi;v_0)$ only depends on $v_0[\phi]$ which is the directional derivative of $\phi$ along the timelike vector $v_0$. The spatial components of the gradient of  $\phi$ are absent in $\cK(\phi;v_0)$ though the scalar field $\phi$ is a function of all of spacetime coordinates. Thus from a classical point of view the action is not usable because the corresponding field equations are insufficient to give  a unique solution for each initial value.

In order to decide whether the creepers are  physically sensible or not we need to explore the corresponding quantum field theory. Path-integral can be used to compute the two-point  functions $D_F(y,y')$ which are the only nontrivial correlation functions in the absence of interactions. It turns out that the quantum field theory is quite reasonable at least in the absence of self-interactions and coupling to the ordinary matter field: Suppose that   $\md$  has a number of   linearly independent nowhere-zero volume-preserving vector fields such that they commute with each other and their  divergence is zero. Assume that we foliate  $\md$ by hypersurfaces generated by these vector fields \cite{Foliat}. Then the path-integral implies that the one-particle states of the creeper $\phi$ are  confined to these hypersurfaces, i.e. the correlation function $D_F(y,y')$ is not zero only if  $y$ and $y'$ are located on the same hypersurface. That is why we name these scalar fields    ``creepers'', though classically they are not confined to any hypersurface.

In the following sections we study these features of the creepers in detail. We start by the definition of the scalar creepers in Minkowski spacetime in section \ref{section-Minkowski} and use the path-integral to compute the correlation function. This computation paves the way to compute the correlation function in nonstationary curved spacetime in section \ref{section-quantization} where the classical creepers are coupled to gravity and their classical field configurations  serve as  a vacuum state corresponding to  a bare cosmological constant.

\section{Scalar creepers in Minkowski spacetime}\label{section-Minkowski}

 In this section we define the scalar  creepers in $\mink$, the $D$ dimensional Minkowski space whose metric  is $\eta={\rm diag}(-1,1,\cdots,1)$,   by giving their classical action. For quantization, we compute the propagator of non-interacting creepers by path integral and show that the propagating modes are localized on $d\le D$ dimensional subspaces.

 The action of $d$ dimensional scalar creepers  in $\mink$ is obtained by removing  $(D-d)$ terms corresponding to directions perpendicular  to a $d$ dimensional hypersurface from the kinetic term of the action of ordinary scalar field theory. It is given by
    \be
    \cS=-\int d^D x \left(\sum_{a,b=0}^{d-1}\frac{1}{2}\eta^{ab}\frac{\p\phi}{\p x^a}\frac{\p\phi}{\p x^b}+V(\phi)\right),
    \label{action-mink}
    \ee
 where
    \be
    V(\phi)=\frac{1}{2}m^2\phi^2+V_{\rm int}(\phi),
    \label{Vint}
    \ee
 $m^2$ is a constant,  and $V_{\rm int}(\phi)$ gives self-interaction. Although  we have dropped kinetic terms corresponding to directions perpendicular  to the hypersurface, we are still considering the scalar field as a function of all spacetime coordinates, i.e.,
    \be
    \phi=\phi(x_\shortparallel, x_\perp),
    \ee
 where $x_\shortparallel^a:=x^a$ for $a=0,\cdots,d-1$, and  $x_\perp^a:=x^a$ for $a=d,\cdots,D-1$.

 Henceforth we study the non-interacting fields and set $V_{\rm int}(\phi)=0$.  The classical field equation is given by ${\delta \cS}/{\delta\phi}=0$, in which,
    \be
    \frac{\delta \cS}{\delta\phi}= \left(\Box^{(d)}-m^2\right)\phi,
    \label{eom-Mink}
    \ee
and
    \be
    \Box^{(d)}:=\sum_{a,b=0}^{d-1}\eta^{ab}\frac{\p^2}{\p x^a\p x^b}.
    \label{box}
    \ee

For $d=D$, Eq.\eqref{action-mink} is the ordinary scalar field theory in Minkowski spacetime. For $d<D$, the classical field equation $\left(\Box^{(d)}-m^2\right)\phi=0$ is not deterministic if not meaningless altogether because it is silent about the behaviour of the classical field in directions $x_\perp$ perpendicular to the hypersurface.
But the classical fields do not participate in particle physics. The particle interpretation of physical states comes from quantum fields whose correlation function is given by  the  path integral \cite{Peskin-book},
    \be
    D_F(x-x'):=\cZ^{-1}\int  \cD\phi\, e^{{\rm i}\cS}\phi(x)\phi(x'),
    \label{1st}
    \ee
where $\cZ:=\int \cD\phi\, e^{{\rm i}\cS}$ is the partition function. Eq.\eqref{eom-Mink}
 implies that
    \be
     \left(\Box^{(d)}-m^2\right)D_F(x-x')=-{\rm i}\cZ^{-1}\int  \cD\phi \frac{\delta e^{{\rm i}\cS}}{\delta \phi(x)}\phi(x').
    \ee
Since ${\delta\phi(x')}/{\delta\phi(x)}=\delta^D(x-x')$, where $\delta^D$ denotes  Dirac delta-function in $D$-dimensions,  integration by parts gives
    \be
    \left(\Box^{(d)}-m^2\right) D_F(x-x')={\rm i}\delta^D(x-x'),
    \label{DF-Mink}
    \ee
whose solution is
\be
    D_F(x-x')= D_F^{(d)}(x_\shortparallel-x'_\shortparallel)\delta^{D-d}(x_\perp-x'_\perp),
    \label{DF-solution-Mink}
    \ee
where  $D_F^{(d)}(x_\shortparallel-x'_\shortparallel)$ denotes the celebrated Feynman propagator in $d$-dimensional Minkowski spacetime \cite{Peskin-book}.

Eq.\eqref{DF-solution-Mink} shows that the path-integral and the correlation function \eqref{1st} are well-defined although the action \eqref{action-mink} is not classically viable since it does not include kinetic terms corresponding to directions normal to the $d$ dimensional hypersurface.  So the action \eqref{action-mink} together with the path-integral \eqref{1st}  describe scalar fields whose one-particle states are delta-function localized on a $d$ dimensional subspace.

\section{Scalar creepers in general spacetimes}\label{section-curved}
In this section we define scalar creepers in general spacetimes and show that, in addition to their ability to stick to lower dimensional hypersurfaces, they hold   a plausible notion of vacuum state and one-particle states in general nonstationary curved spacetimes, in contrast to  ordinary scalar field theories \cite{Wald-Book}.
Thus, before discussing the scalar creepers we review the ordinary scalar field theory  briefly in order to identify the main obstacles in computing their  propagator via path integral in general  spacetimes.

To see the problem with ordinary scalar field theory, consider the minimally coupled massless scalar field theory in $\md$  whose action is  given by
    \be
    \bS:=-\frac{1}{2}\int d^D y\,\fe\, g^{\mu\nu}\p_\mu\varphi\p_\nu\varphi,
    \label{minimal-1}
    \ee
where $\fe:=\sqrt{\left|\det g\right|}$ and $\p_\mu:=\frac{\p}{\p y^\mu}$.  By introducing the tetrad ${e^\mu}_a$ satisfying $\eta^{ab}{e^\mu}_a{e^\nu}_b=g^{\mu\nu}$ and the vector fields ${e_a}:={e^\mu}_a\p_\mu$, the action \eqref{minimal-1} can be written as
    \be
     \bS:=-\frac{1}{2}\int d^D y\,\fe\,\eta^{ab}e_a[\varphi]e_b[\varphi],
    \label{minimal-2}
    \ee
 where $e_a[\varphi]:={e^\mu}_a\p_\mu\varphi$ is diffeomorphism invariant.

Although the action \eqref{minimal-2}  is similar to Eq.\eqref{action-mink} we cannot, in general,  compute the corresponding path integral and obtain the correlation function explicitly. The difficulty can be seen from the expression
    \be
    \label{minimal-eom}
    \frac{\delta\bS}{\delta\varphi}=\fe\eta^{ab}\Big({e_a}{e_b}+(\nabla_\mu {e_a}^\mu){e_b}\Big)\varphi,
    \ee
where  $\nabla_\mu$ denotes the Levi-Civita connection, and we have used the identity
    \be
    \nabla_\mu {v}^\mu=\fe^{-1}\p_\mu (\fe\, {v}^\mu).
    \label{indentity}
    \ee
Comparing Eq.\eqref{minimal-eom} with equations \eqref{eom-Mink} and \eqref{box} reveals the roots of  the difficulty: the vector fields ${e_a}$ are not necessarily   divergence free and they do not commute with each other in general.

In order to define the creepers in general nonstationary curved spacetimes we  need to replace the tetrad $e_a$ in Eq.\eqref{minimal-2}  with a set of nowhere-zero vector fields $v_a:={v_a}^\mu\partial_\mu$ such that
    \be
    [{v_a},{v_b}]\phi:=\left({v_a}^\mu \nabla_\mu{v_b}^\nu-{v_b}^\mu\nabla_\mu {v_a}^\nu\right)\p_\nu\phi=0,
    \label{prop-1}
    \ee
and
    \be
    \nabla_\mu {v_a}^\mu=0.
    \label{prop-2}
    \ee

In subsection \ref{section-geometry} we show that the vector fields $v_a$ exist locally though they do not exist globally.  We introduce the scalar creepers  in subsection \ref{section-action} and define their one-particle states   in subsection \ref{section-quantization}.
\subsection{Geometry}\label{section-geometry}
A straightforward approach to obtain the vector fields $v_a$ satisfying equations \eqref{prop-1} and \eqref{prop-2}  is to work with coordinate systems $x^\mu$ used in unimodular gravity \cite{Anderson,Padilla:2014yea} in which $\fe=1$. In these coordinates  ${v_a}^\mu={\boldsymbol{\delta}_a}^\mu$ where $\boldsymbol{\delta}$ denotes  the Kronecker delta, i.e.,
    \be
    {v_a}:={v_a}^\mu\p_\mu=\frac{\p}{\p x^a}.
    \ee
To confirm this proposition we only need to use the identities
    \begin{align}
    \label{prop-1-formula}
    &{v_a}^\mu\nabla_\mu {v_b}^\nu-{v_b}^\mu\nabla_\mu {v_a}^\nu={v_a}^\mu\p_\mu {v_b}^\nu-{v_b}^\mu\p_\mu {v_a}^\nu,\\
    \label{prop-2-formula}
    &\nabla_\mu {v}^\mu=\fe^{-1}\p_\mu (\fe\, {v}^\mu).
    \end{align}

Henceforth  by $x$-coordinates we mean a coordinate system in which $\fe=1$. In order to find such coordinates explicitly  we rearrange the coordinates $y^\mu$  as $(y^0,\vec y)$  and suppose that
    \begin{align}
    &{x^0}(y^0,\vec y):=\int_{y^0_{\rm ref}}^{ y^0}dz\, \fe(z,\vec y),&{\vec x}:={\vec y},
    \label{x-minus}
    \end{align}
where  $y_{\rm ref}^0=y_{\rm ref}^0(\vec y)$ is a reference point. One verifies that $d^D x=\fe\,d^Dy$ which implies that the determinant of the metric in the $x$-coordinates equals -1. Furthermore $\frac{\p}{\p x^0}$ is timelike as long as $\p_0$ is timelike because
    \begin{align}
    \label{timelike}
    &\frac{\p}{\p x^0}=\fe ^{-1}\p_0,\\
    \label{spacelike}
    &\frac{\p}{\p  x^i}= -\fe^{-1}\left(\p_i x^0\right)\p_0+\p_i, &i=1,\cdots,D-1.
     \end{align}

As an example consider the four dimensional Schwarzschild spacetime in Kruskal-Szekeres coordinates
    \be
    ds^2=4r^{-1} e^{-r}du dv- r^2d\Omega^2,
    \ee
where $d\Omega^2:=\frac{d \fc^2}{1-\fc^2}+(1-\fc^2)d\varphi^2$, $\fc:=\cos\vartheta$,\footnote{We are replacing $\vartheta$ with $\fc$ in order to make $\fe$ independent of $\vartheta$.} and  $r=1+W$, in which $W:=W(-uve^{-1})$ denotes the Lambert $W$ function. Thus, $uv=e^{r}(1-r)$.
Eq.\eqref{x-minus} (with $u$ playing the role of $y^0$) reads
    \be
    x^{0}=-\frac{2W}{3v}(W^2+3W+3),
    \label{x-minus-Sch}
    \ee
where we have supposed that  $y_{\rm ref}^0(\vec y)=0$, i.e., the reference point is located on  the event horizon $r=1$. $W(-uve^{-1})$ is real-valued for $uv\le 1$ or equivalently for $r\ge 0$.

\subsubsection{Closed FRW universe}
 So far we have shown that in a $D$ dimensional spacetime  there always exist $D$ vector fields satisfying equations \eqref{prop-1} and \eqref{prop-2} locally. For defining creepers we need to know how many of them  exist globally. The existence of nowhere-zero vector fields is a question in homotopy theory.  Take $n$  dimensional spheres $S^n$ for instance. We know that for $n=1,3$ there exist exactly $n$ nowhere-zero vector fields while there is no such vector fields on $S^2$ \cite{Adams}. In this subsection we show that in a $D=4$ dimensional FRW universe with closed spatial sections only two nowhere-zero vector fields satisfying equations \eqref{prop-1} and \eqref{prop-2} coexist globally.

 Suppose that
    \be
    ds^2=-dt^2+\omega(t)^2 ds^2_\Sigma,
    \ee
 where $\Sigma$, whose line element $ds^2_\Sigma$ is $t$-independent, is diffeomorphic to a three sphere $S^3$ \cite{Hawking:1973uf}. The $x^0$ coordinate is given by Eq.\eqref{x-minus}
    \be
    x^0=\int^{t} dt'\omega(t')^{3},
    \ee
and consequently $v_0:=\omega(t)^{-3}\p_t$. The vector field $v_0$ exists and its divergence is  zero except for the beginning and the  end when $\omega(t)=0$.

Now we focus on the space section $\Sigma$, which we model by $S^3$,
     \be
     S^3:=\left\{\left(X^1,X^2,X^3,X^4\right)\Big\vert\sum_{i=1}^{4}\left(X^i\right)^2=1\right\},
     \ee
embedded in $(\mathbb{R}^4,\boldsymbol{\delta})$.  $S^3$ is parallelizable, i.e., there exist  three independent vector fields on $S^3$,
    \begin{align}
    &v_1:=(-X^2,X^1,-X^4,X^3),\\
    &v_2:=(-X^3,X^4,X^1,-X^2),\\
    &v_3:=(-X^4,-X^3,X^2,X^1).
    \end{align}
The divergence of $v_i$ is zero but
    \begin{align}
    &[v_i,v_j]=-2\sum_{k=1}^{3}\epsilon_{ijk} v_k,&i,j=1,2,3.
    \end{align}
 Thus, in an FRW universe with closed spatial sections there exist at most two vector fields, e.g.,  $v_0$ and $v_1$ satisfying equations \eqref{prop-1} and \eqref{prop-2} simultaneously.  In the coordinate system $(\theta,\phi_1,\phi_2)$ given by
    \begin{align}
    &X_{2I-1}=r_I\cos\phi_I,\\
    &X_{2I}=r_I\sin\phi_I,
    \end{align}
 where $I=1,2$, and $r_1:=\cos\theta$ and $r_2:=\sin\theta$, we have
    \be
    ds^2_\Sigma=d\theta^2+\cos^2\theta d\phi_1^2+\sin^2\theta d\phi_2^2,
    \ee
 and
    \be
    v_1=\frac{\p}{\p\phi_1} +\frac{\p}{\p\phi_2}.
    \ee

\subsection{Action}\label{section-action}
In  a $D$ dimensional spacetime $\md$  with $d_*\le D$ linearly independent nowhere-zero  vector fields $v_a$ satisfying equations \eqref{prop-1} and \eqref{prop-2}, we give the $d\le d_*$ dimensional creepers' action  by
    \be
    \cS:=\int d^Dy\, \fe\cL(\phi;v_a),
    \label{action}
    \ee
 in which the creeper $\phi=\phi(y^0,\cdots,y^{D-1})$ is a scalar field, $\fe:=\sqrt{\left|\det g\right|}$,
    \be
    \cL(\phi;v_a):=-\frac{1}{2}\sum_{a,b=0}^{d-1}\eta^{ab}v_a[\phi]
    v_b[\phi]-V(\phi),
    \label{main-new}
    \ee
 and the potential $V(\phi)$ is given by Eq.\eqref{Vint}. Following the definition of vector fields, $v_a[\phi]:={v_a}^\mu\partial_\mu\phi$ is a scalar,  therefore the Lagrangian density $\cL(\phi;v_a)$ is also a scalar  though it is independent of the spacetime metric $g$. Thus $\cS$ is diffeomorphism invariant off-shell.

Using equations  \eqref{prop-2} and  \eqref{prop-2-formula} one verifies that the classical equation of motion is given by $\frac{\delta \cS}{\delta\phi}=0$, where
    \be
    \label{eom-GCST-original}
    \frac{\delta \cS}{\delta\phi}=\fe \left(\Box^{(d)}\phi-\frac{\p V(\phi)}{\p \phi}\right),
    \ee
and
    \be
    \Box^{(d)}\phi:=\eta^{ab}v_a\left[v_b[\phi]\right].
    \label{box-curve}
    \ee
Later in subsection \ref{section-quantization} we use this result to compute the correlation function by path integral.
\subsubsection{Stress tensor}\label{section-stress}
Now we couple the scalar creepers to Einstein's gravity. For this purpose we first recall  the definition of the stress tensor and the consequences of the  off-shell diffeomorphism invariance in ordinary matter field theory.

In a theory whose action is $\bS=\int d^Dy\,\fe\bL$ the stress tensor is defined by
    \be
    \bT_{\mu\nu}:=-\frac{2}{\fe}\frac{\delta\bS}{\delta g^{\mu\nu}}.
    \ee
An infinitesimal diffeomorphism is given by the map $y\to y_\xi$ such that $\delta_\xi y:={y_\xi}-y=\xi(y)$. We suppose that  the vector field $\xi(y)$ is zero on the boundary and  drop all the boundary terms subsequently.  The corresponding  variation of a scalar $\phi$, a vector $v$, the one-form  $d\phi$, the tensor  $g$ and the volume element $\fe$ are given by the Lie derivatives
    \begin{align}
    \label{xi-phi}
    &\delta_\xi\phi(y)=-\xi^\mu\p_\mu\phi(y),\\
    \label{xi-d-phi}
    &\delta_\xi\left(\p_\mu\phi(y)\right)=-\p_\mu\left(\xi^\nu\p_\nu\phi(y)\right)=
    \p_\mu\left(\delta_\xi\phi\right),\\
    \label{xi-v}
    &\delta_\xi v^\mu=-\xi^\nu\nabla_\nu v^\mu+v^\nu\nabla_\nu\xi^\mu,\\
    \label{xi-g}
    &\delta_\xi g_{\mu\nu}=-\nabla_\nu\xi_\mu-\nabla_\mu\xi_\nu,\\
    \label{xi-e}
    &\delta_\xi \fe=-\nabla_\mu\xi^\mu.
    \end{align}
Since $\bL$ is a scalar, similarly to Eq.\eqref{xi-phi} we have $\delta_\xi\bL=-\xi^\mu\p_\mu\bL$ which  together with Eq.\eqref{xi-e} results in the off-shell diffeomorphism invariance  $\delta_\xi\bS=0$.

For ordinary matter fields we have
    \be
    \delta_\xi \bS= \int d^Dy \frac{\delta \cS}{\delta g_{\mu\nu}}\delta_\xi g_{\mu\nu}+\int d^Dy\,\fe\left[\frac{\p \bL}{\p\left(\p_\mu\phi\right)}\delta_\xi\left(\p_\mu\phi(y)\right)+\frac{\p \bL}{\p\phi}\delta_\xi\phi\right].
    \label{diff-S}
    \ee
Using equations \eqref{xi-phi} and \eqref{xi-d-phi} together with the classical field equation
    \be
    \frac{\p \bL}{\p\phi}=\nabla_\mu\frac{\p \bL}{\p\left(\p_\mu\phi\right)},
    \ee
one verifies that the second term on the right hand side of Eq.\eqref{diff-S} vanishes. Furthermore, by using Eq.\eqref{xi-g} and integration by parts we obtain
    \be
     \int d^Dy \frac{\delta \cS}{\delta g_{\mu\nu}}\delta_\xi g_{\mu\nu}=\int d^D y \,\fe\,\xi_\nu\nabla_\mu \bT^{\mu\nu}.
    \ee
Thus in ordinary scalar theory, the off-shell diffeomorphism invariance together with the classical field equation imply that the stress tensor is conserved on-shell, i.e.,
    \be
     \left.\nabla_\mu \bT^{\mu\nu}\right\vert_{\rm on-shell}=0.
    \label{conservation-matter}
    \ee

Similarly,  the Einstein-Hilbert action $S_{\rm EH}$ (in the metric formulation) depends on the spacetime metric only and
    \be
    \delta_\xi S_{\rm EH}= \int d^Dy \frac{\delta S_{\rm EH}}{\delta g_{\mu\nu}}\delta_\xi g_{\mu\nu}.
    \ee
Defining  the  Einstein tensor by
    \be
    G_{\mu\nu}:=\frac{2}{\fe}\frac{\delta S_{\rm EH}}{\delta g^{\mu\nu}},
    \ee
 and using Eq.\eqref{xi-g} and $\delta_\xi S_{\rm EH}=0$, we conclude that $\nabla_\mu G^{\mu\nu}=0$. On the other hand,  Einstein's field equation given by
    \be
    \frac{\delta}{\delta g_{\mu\nu}}\left(S_{\rm EH}+\bS\right)=0,
    \ee
 reads $    G_{\mu\nu}=\bT_{\mu\nu}$.
Thus, Eq.\eqref{conservation-matter} shows that ordinary scalar theories whose actions are diffeomorphism invariant off-shell can be coupled to Einstein's gravity on-shell.

In the case of scalar creepers,  $\cL(\phi;v_a)$ is independent of the metric $g_{\mu\nu}$ and Eq.\eqref{action} together with the identity $\delta \fe=\frac{1}{2}\fe g^{\mu\nu}\delta g_{\mu\nu}$ imply that the  creepers' stress tensor is given by
\be
    T_{\mu\nu}:=-\frac{2}{\fe}\frac{\delta\cS}{\delta g^{\mu\nu}}=\cL(\phi;v_a)\,g_{\mu\nu},
    \label{stress}
    \ee
which resembles a perfect fluid with equation of state $w=-1$. Coupling the scalar creepers to Einstein's gravity results in the field equation
    \be
    \frac{\delta}{\delta g_{\mu\nu}}\left(S_{\rm EH}+\cS\right)=0,
    \ee
which gives $G_{\mu\nu}=T_{\mu\nu}$. Consistency of this equation with the identity $\nabla_\mu G^{\mu\nu}=0$ requires that
    \be
     \left.\nabla_\mu T^{\mu\nu}\right\vert_{\rm on-shell}=0,
    \label{conservation-creeper}
    \ee
Therefore a consistent coupling to gravity requires
    \be
     \left.\cL(\phi;v_a)\right\vert_{\rm on-shell}=\mbox{constant}.
    \label{L-is-constant}
    \ee
In other words,  the scalar creepers couple to gravity similarly to a cosmological constant term. We study this phenomenon  in section \ref{section-cosmology}.

We should not have expected to obtain Eq.\eqref{conservation-creeper} as a direct consequence of the off-shell diffeomorphism invariance of $\cS$ and the classical field equation \eqref{eom-GCST-original} similarly to Eq.\eqref{conservation-matter}, because there exist solutions to the classical field equation for which the on-shell value of $\cL(\phi;v_a)$ is not constant. In fact  we can not obtain Eq.\eqref{conservation-creeper} by following those steps that led us to  Eq.\eqref{conservation-matter} although  $\cL(\phi;v_a)$ is a scalar and consequently  $\delta_\xi \cS=0$ off-shell. This can be verified by noting that
    \be
    \delta_\xi \cS= \int d^Dy \frac{\delta \cS}{\delta g_{\mu\nu}}\delta_\xi g_{\mu\nu}+\int d^Dy\frac{\delta \cS}{\delta\phi}\delta_\xi\phi
    +\int d^Dy\,\fe\eta^{ab}\left(v_a[\phi]\right)\left([\xi,v_b]\phi\right),
    \label{trial}
    \ee
where  we have used equations \eqref{xi-v} to obtain the last term which is new compared to Eq.\eqref{diff-S}.  By integration by parts and using equations  \eqref{prop-2}, \eqref{eom-GCST-original}, \eqref{box-curve} and \eqref{xi-phi} we obtain
    \be
    \int d^Dy\,\fe\eta^{ab}\left(v_a[\phi]\right)\left([\xi,v_b]\phi\right)=\int d^Dy\,\fe\delta_\xi\cL(\phi;v_a)-\int d^Dy\,\fe\frac{\delta \cS}{\delta\phi}\delta_\xi\phi.
    \label{trial-2}
    \ee
So the third term in  Eq.\eqref{trial} cancels the $\delta\cS/\delta\phi$ term therein and by inserting Eq.\eqref{trial-2} in Eq.\eqref{trial} we obtain
    \be
    \delta_\xi \cS= \int d^Dy \frac{\delta \cS}{\delta g_{\mu\nu}}\delta_\xi g_{\mu\nu}+\int d^Dy\,\fe\delta_\xi\cL(\phi;v_a).
    \label{trial-1}
    \ee
 Thus there is no contribution from the classical field equation. After using equations \eqref{stress}, \eqref{xi-g} and \eqref{xi-phi} in Eq.\eqref{trial-1} we simply end up with $\delta_\xi\cS=0$.

In summary, a scalar creeper on a $d\le D$ dimensional hypersurface is defined by equations \eqref{action} and \eqref{main-new}.  The creepers are perfect fluids with equation of state $w=-1$ and consequently they do not correspond to ordinary matter field. In order to couple  the scalar creepers to gravity  their Lagrangian should be constant on-shell.

\subsection{Quantization}\label{section-quantization}
The propagator of the $d$ dimensional scalar creepers in a $\md$ can be computed by using the  classical action in the path integral. For $V_{\rm int}(\phi)=0$ the path integral
    \be
    D_F(y,y'):=\cZ^{-1}\int  \cD\phi\, e^{{\rm i}\cS}\phi(y)\phi(y'),
    \ee
together with Eq.\eqref {eom-GCST-original}  give
    \be
    \label{DF-GCST-original}
    \left(\Box^{(d)}-m^2\right)\!\! D_F(y,y')=-\frac{{\rm i}}{\cZ}\int  \cD\phi \,\fe(y)^{-1} \frac{\delta e^{{\rm i}\cS}}{\delta \phi(y)}\phi(y')=\frac{{\rm i}}{\fe(y)}\delta^D(y-y'),
    \ee
in spite of the fact that we have not defined the path integral, especially the temporal (Feynman) boundary conditions yet. We have only assumed that the path integral exists and the integration by parts is applicable. To compute $D_F(y,y')$ we use the $x$-coordinates.

\subsubsection{The $x$-coordinates revisited}\label{section-revisit}
    In Eq.\eqref{x-minus} we observed that there is a local coordinate system  in which the volume element $\fe=1$. By locality we mean that  this coordinate system is valid on the $y$-coordinate neighborhood as indicated by  Eq.\eqref{x-minus}. Therefore as long as  the $y$-coordinates do not necessarily cover the whole spacetime there is no guarantee that the corresponding  $x$-coordinates can be defined everywhere.\footnote{There is a sharp distinction between the $x$-coordinates and the local Lorentz frame in the vicinity of a spacetime event $\cP$ defined by $g_{\mu\nu}(\cP)=\eta_{\mu\nu}$ and $\p_\mu g_{\nu\rho}(\cP)=0$ whose domain is limited by the curvature.} On the other hand the  nowhere-zero vector fields $v_i$, $i=1,\cdots,\tilde{d}\le d_*\le D$ satisfying the condition \eqref{prop-1} are ``integrable,'' so we can use them to foliate the spacetime \cite{Foliat} and presumably define  $d\le\tilde{d}$ dimensional scalar creepers.

 Since  the foliation $\{\sL_\alpha\}$ generated by the vector fields $v_i$ is a global structure, it is natural to use it to define the coordinate system in the first place. So we look for a local coordinate system $x^\mu$, $\mu=0,\cdots,D-1$ such that
 \begin{enumerate}
 \item the volume element in the $x$-coordinates equals 1,
 \item ${v_i}^\mu=\boldsymbol{\delta}_i^\mu$.
 \end{enumerate}
  These properties are also essential for our method for solving Eq.\eqref{DF-GCST-original} and quantizing the scalar creepers, so in  the following we  describe a construction of the $x$-coordinates satisfying the above mentioned properties.

  In appendix \ref{section-appendixA} we show that as a consequence of the property \eqref{prop-1} every point in $\cM$ has a neighborhood $U$ and a system of local coordinates  $y^\mu$, $\mu=0,\cdots,D$  such that the components of $U\cap \sL_\alpha$  are described by the equations
    \begin{align}
     &y^\mu=\mbox{constant} &\mbox{for}\ \  \mu\not\in{\fI}:=\{1,\cdots,\tilde{d}\},
    \end{align}
and
  \begin{align}
  &v_i=\p_i &\mbox{for}\ \ \ i\in\fI.
 \label{v-xf}
  \end{align}
  In this coordinate neighborhood we define the $x$-coordinates by Eq.\eqref{x-minus}. Thus  the volume element in the $x$-coordinates equals 1 and the first condition mentioned above is fulfilled in this construction. To address the second condition, suppose that
    \begin{align}
     &v_i[\fe]=\p_i\fe =0&\mbox{for}\ \ i\in\cI\subset\fI.
    \end{align}
  If $\cI$ is nonempty, i.e. if some of the vector fields $v_i$ are volume-preserving,  we can relabel the vector fields $v_i$ such that $\cI=\{1,\cdots,d_I\le\tilde{d}\}$. So Eq.\eqref{x-minus} implies that $\p_ix^0=0$ for $i\in\cI$ which  together with Eq.\eqref{spacelike} and Eq.\eqref{v-xf}  give
    \begin{align}
     &v_i=\frac{\p}{\p  x^i}&\mbox{for}\ \ i\in\cI.
    \end{align}
  So in this construction of the $x$-coordinates the second conditions is fulfilled  only for the volume-preserving  vector field $v_i$, $i\in\cI$.  Consequently for quantization we focus on $d\le d_I$ dimensional creepers whose  action \eqref{action} in the  $x$-coordinates reads
    \begin{align}
    &\cS=-\int d^Dx\left(\frac{1}{2}\sum_{a,b=0}^{d}\eta^{ab}\frac{\p\phi}{\p x^a}\frac{\p\phi}{\p x^b}+V(\phi)\right),&d\le d_\cI,
    \label{action-GCST}
    \end{align}
 resembling   the action \eqref{action-mink}.

\subsubsection{The two-point function}
To compute $D_F(y,y')$ by solving Eq.\eqref{DF-GCST-original} we note that the creepers' action in  the $x$-coordinate system  given by Eq.\eqref{action-GCST}  is identical to Eq.\eqref{action-mink},  the operator $\Box^{(d)}$ defined in Eq.\eqref{box-curve} is given by
    \begin{align}
    &\Box^{(d)}=\sum_{a,b=0}^{d-1}\eta^{ab}\frac{\p^2}{\p x^a\p x^b},&d\le d_\cI,
     \end{align}
similarly to Eq.\eqref{box}, and $\frac{1}{\fe}\delta^D(y-y')=\delta^D(x-x')$. Therefore Eq.\eqref{DF-GCST-original} in the $x$-coordinates is equivalent to Eq.\eqref{DF-Mink}. Consequently the correlation function is given by Eq.\eqref{DF-solution-Mink},  and  the corresponding particle description is also applicable here, though the $d\le D$ dimensional subspace is embedded in a nonstationary spacetime. Precisely we can  consider an auxiliary  action
    \be
    \label{2D-action}
    \cS=-\frac{1}{2}\int d^dx_\shortparallel \left(\sum_{a,b=0}^{d-1}\eta^{ab}\frac{\p\Phi(x_\shortparallel)}{\p x^a}\frac{\p\Phi(x_\shortparallel)}{\p x^a}+m^2\Phi(x_\shortparallel)^2\right),
    \ee
whose second quantization  indicates that  there exist a vacuum state $\left|0\right\rangle$ with respect to the ``time coordinate'' $x^0$ (cf. Eq.\eqref{timelike}) and a set of creation and annihilation operators $a^\dag({\bf p})$ and $a({\bf p})$ with commutation relations
    \begin{align}
      &[a({\bf p}),a({\bf q})]=0,\\
      & [a({\bf p}),a^\dag({\bf p}')]=(2\pi)^{d-1}\delta^{d-1}({\bf p}-{\bf p}'),\\&[a^\dag({\bf p}),a^\dag({\bf p}')]=0,
    \end{align}
such that $a({\bf p})\left\vert0\right\rangle=0$ \cite{Peskin-book}. The field operator is given by
%\begin{widetext}
    \be
    \Phi_\shortparallel(x_\shortparallel):=
    \int_{-\infty}^{\infty}\frac{d^{d-1}{\bf p}}{(2\pi)^{d-1}\sqrt{2E({\bf p})}}
    \left[a({\bf p})e^{-{\rm i}p\cdot x_\shortparallel} +a^\dag({\bf p})e^{{\rm i}p\cdot x_\shortparallel}\right],
    \label{2nd-quantized}
    \ee
%\end{widetext}
where $x_\shortparallel:=(x^0,{\bf x})$, $p:=(E({\bf p}),{\bf p})$, $E({\bf p}):=\sqrt{{\bf p}\cdot{\bf p}+m^2}$,  $p\cdot x_\shortparallel:=E(p)x^0-{\bf p}\cdot {\bf x}$, and the inner product
    \be
    {\bf u}\cdot{\bf w}:=\sum_{a=1}^{d-1}u^a w^a.
     \ee
We postulate, and define the  path integral accordingly, that $D_F^{(d)}$ in Eq.\eqref{DF-solution-Mink} equals the corresponding Feynman propagator given by\footnote{ $\theta(x)$ denotes the Heaviside step function. $\theta(x)$ equals 1 and 0, for $x>0$ and $x<0$ respectively.}
    \be\label{Feynman}
    \theta(x^0-{x'}^0)\left\langle0\Big\vert\Phi_\shortparallel(x_\shortparallel)\Phi_\shortparallel(x'_\shortparallel)\Big\vert0\right\rangle
    +x\leftrightarrow x',
    \ee
which gives  the amplitude for particles with ``positive frequency'' to propagate from $x_\shortparallel'$ to $x_\shortparallel$ and from $x_\shortparallel$ to $x'_\shortparallel$  for $x^0>{x'}^0$ and $x^0<{x'}^0$ respectively \cite{Peskin-book}.
Consequently, we have a  notion of vacuum state, creation and annihilation operators and ``time-ordering'' of $n$-point functions  with respect to the  ``time-coordinate'' $x^0$, although  Eq.\eqref{timelike} demonstrates that the vector field $v_0$ is not necessarily timelike everywhere.

In particular, for $d=2$ and $m=0$, Eq.\eqref{action-GCST} reads
    \be
    \cS=2\int d^Dx \,\p_+\phi\p_-\phi,
    \ee
in which $\p_\pm:=\frac{1}{2}\left(\frac{\p}{\p x^0}\pm \frac{\p}{\p x^1}\right)$. This theory  can be interpreted as  a $c=1$ conformal field theory \cite{Ginsparg} embedded in $\md$.
\section{Application to cosmology}\label{section-cosmology}
In section \ref{section-curved} we defined the scalar creepers in $\md$ and observed that they have well-defined one-particle states localized to $d$ dimensional hypersurfaces of the bulk. In this section we couple the scalar creepers to gravity and investigate their contribution to the dark energy.

In subsection \ref{section-stress} we showed that the scalar creepers can be coupled to gravity consistently  if
    \be
    \left.\nabla_\mu T^{\mu\nu}\right|_{\rm{on-shell}}=0.
    \ee
This condition together with Eq.\eqref{stress} implies that the creepers can be coupled to Einstein's gravity only if $\cL(\phi;v_a)$ is constant on-shell. In this way, $T_{\mu\nu}$ resembles the
 bare cosmological constant term $\lambda_B$ in the Einstein's field equation suggesting  that
    \be
    {\lambda_B}=-{8\pi G}\left.\cL(\phi;v_a)\right|_{\rm{on-shell}}.
    \label{lambda-b}
    \ee
 For the classical solution $\phi=\bar\phi={\rm constant}$, such that
    \be
    \left.\frac{dV(\phi)}{d\phi}\right|_{\phi=\bar{\phi}}=0,
    \label{minimum-1}
    \ee
we have
    \be
    \left.\cL\right|_{\rm on-shell}=-V(\bar{\phi}).
    \label{minimum-2}
    \ee
Similarly to  ordinary scalar field theories, such solutions are not interesting because they need a fine-tuned potential according to the requirements \eqref{lambda-b}, \eqref{minimum-1} and \eqref{minimum-2} \cite{Joyce}. Thus, we dismiss such solutions and suppose that the creepers are not  constant on-shell, though we seek  classical solutions such that  $\cL$ is  constant on-shell, corresponding to a bare cosmological constant. So we demand
    \begin{align}
    &\phi_{\rm on-shell}\neq {\rm constant},&
    \cL_{\rm on-shell}= {\rm constant}.
    \label{requirements}
    \end{align}

In the following we study  $d=1$ and $d\ge 1$ creepers separately. In the one dimensional case we verify that the requirements \eqref{requirements} imply that $V(\phi)$ is necessarily zero and  $\lambda_B< 0$. For $d\ge 1$ we confine our study to free creepers with $V(\phi)=0$ and  identify the parameter space of classical solutions satisfying the requirements \eqref{requirements}. We discover that for $d=2$ the parameter space enjoys an $O(1,1)$ symmetry which enhances to $\mathbb{Z}_2\times{\rm Diff}(\mathbb{R}^1)$ at $\lambda_B=0$, while for $d>2$  the symmetry is  $O(d-1,1)$,  $O(d-1)\times {\rm Diff}(\mathbb{R}^1)$  and $O(d-1,1)\times O(d-2)\times {\rm Diff}(\mathbb{R}^1)$ respectively for  $\lambda_B<0$,  $\lambda_B=0$ and, $\lambda_B>0$.
\subsection{$d=1$ dimensional scalar creepers}
One-dimensional creepers exist in time-orientable spacetimes, cf. section \ref{section-Prologue}.\footnote{Here we are concerned with the classical theory. For quantization we can apply the method of section \ref{section-quantization}  in stationary spacetimes and in FRW universes.}
 For $d=1$ the field equation \eqref{eom-GCST-original} together with Eq.\eqref{lambda-b} imply that
    \begin{align}
    \label{d=1-1}
    {\dot{\phi}}^2=\frac{\lambda_0-\lambda_B}{8\pi G},\\
    \label{d=1-2}
    V(\phi)=\frac{\lambda_B+\lambda_0}{8\pi G},
    \end{align}
 where $\lambda_0$ is an integration constant, and
    \be
    \dot\phi:=\frac{\delta\cS}{\delta v_0[\phi]}=v_0[\phi],
    \ee
 is the momentum conjugate to $\phi$. Recalling Eq.\eqref{Vint} we note that there exists no function $V(\phi)$ except for $V(\phi)=0$ satisfying  \eqref{requirements},  \eqref{d=1-1} and \eqref{d=1-2}. Therefore $\lambda_0=-\lambda_B$, and Eq.\eqref{lambda-b} reads,
    \be
    \lambda_B=-4\pi G{\dot\phi}^2<0.
    \label{lambda-momentum}
    \ee

 To estimate the value of $\lambda_B$ in the $D=4$ model, we note that since $V(\phi)=0$ the only energy  scale at hand is $M_{\rm Pl}$. So we can rewrite  the action \eqref{action} and the corresponding classical field equation in terms of the   dimensionless field $\bar\phi:=M_{\rm Pl}^{-1}\phi$  and the dimensionless coordinates $\bar{y}:=M_{\rm Pl}y$ as
    \begin{align}
    &\cS=\frac{1}{2}\int d^4\bar{y}\,\fe {\bar{v}_0}[\bar{\phi}]^2,\\
    &{\bar{v}_0}\left[{\bar{v}_0}[\bar{\phi}]\right]=0,
    \end{align}
 where ${\bar{v}_0}:=M_{\rm Pl}^{-1}{v_0}$. The classical solutions is $\bar\phi=\kappa({\bar{x}}^0-\fx^0)$, where ${\bar{x}}^0:=M_{\rm Pl}x^0$ and  $\fx^0\in\mathbb{R}$ is an integration constant.
So it is natural\footnote{Since  the transition rates  in particle physics are extremely slow compared to $1M_{\rm Pl}$, there should exist  some other physical processes, inevitably in the gravity side, whose rates are of order $1M_{\rm Pl}$. The dimensionless gravitational degrees of freedom are the metric components $g_{\mu\nu}$ and the creeper $\bar\phi$, hence we are suggesting ${v_0}\bar{\phi}$  as the natural candidate.} to assume that $\left|\kappa\right|\sim 1$ and consequently
    \be
    \left|\lambda_B\right|\sim 1M_{\rm Pl}^2.
     \label{assumption}
     \ee
 Thus equation \eqref{lambda-b} satisfies the conditions required in the references \cite{Wang,Unruh}, i.e.,  $\lambda_B<0$ and
    \be
    -\lambda_B \gg \Lambda^2,
    \ee
 where $\Lambda\sim 10^{-14} M_{\rm Pl}$ is the  high energy cutoff for the  ordinary quantum field theory \cite{Carmona}. It is important to note that this result is independent of how we choose $v_0$ and  $\fx^0$.

 This result has two consequences. First of all we can identify $\bar\phi$ with a continuous  and monotonic time which according to Eq.\eqref{lambda-momentum} does not commute with  $\lambda_B$ \cite{Smolin}. Secondly, since  $\left|\kappa\right|\sim 1$ quite naturally, we conclude that the Wang-Unruh approach to the cosmological constant problem provides an anthropic explanation of the hierarchy problem: since $\left|\lambda_B\right|\sim 1M_{\rm PL}^2$ and $\lambda_{\rm eff}\sim 10^{-122}M_{\rm Pl}^2$ we should have $\Lambda \ll 1$ in Planck units \cite{Wang,Unruh}.
\subsection{$d\ge 2$ dimensional scalar creepers}

Suppose  $V(\phi)=0$. In this case the action \eqref{main-new} is invariant under adding an arbitrary constant $\phi_0$ to $\phi$. So in the following we study the equivalence classes of classical solutions defined accordingly, i.e., throughout this subsection, $\phi$ stands for the set of all fields $\tilde{\phi}$ such that $\tilde{\phi}(y)-\phi(y)$ is constant.

 For $d\ge 2$  the classical solutions to the field equation $\Box^{(d)}\phi=0$ satisfying the requirements \eqref{requirements}  are given by
    \be
    \phi_{\rm cl}=\sum_{a=0}^{d-1}\kappa_a x^a+f(z),
    \label{phi-cl-general}
    \ee
in which   $f$ is a  smooth function and
    \be
    z:= \sum_{a=0}^{d-1}\fK_a x^a,
    \ee
such that
    \be
    \begin{array}{ccc}
    \fK\cdot \fK=0, &\fK\cdot \kappa=0,&  \kappa\cdot\kappa=(4\pi G)^{-1}\lambda_B,
    \end{array}
    \label{parameter}
    \ee
 where, for example,
    \begin{align}
    &\fK\cdot\kappa:=\sum_{a,b=0}^{d-1}\eta^{ab}\fK_a \kappa_b.
    \end{align}
In the following we investigate the parameter space of the classical solutions \eqref{phi-cl-general} indicated by equations \eqref{parameter} for $\lambda_B<0$, $\lambda_B=0$ and $\lambda_B>0$ separately.
\subsubsection{Classical solutions corresponding to $\lambda_B<0$}
Since $\kappa\cdot\kappa=(4\pi G)^{-1}\lambda_B$ we can choose $\kappa=(\sqrt{\left|\kappa\cdot\kappa\right|},0,\cdots,0)$, which together with  $\fK\cdot\kappa=0$ imply that $\fK_0=0$. Using this result in $\fK\cdot \fK=0$ we obtain  $\fK=0$. Thus the classical solution is
    \be
    \phi_{\rm cl}=\sum_{a=0}^{d-1}\kappa_a x^a,
    \ee
the parameter space includes only $\kappa_a$'s, and it is symmetric under the action of the group $O(d-1,1)$.
\subsubsection{Classical solutions corresponding to  $\lambda_B= 0$}
In this case $\kappa\cdot\kappa=0$. Therefore in Eq.\eqref{phi-cl-general} we can absorb the term $\sum_{a=0}^{d-1}\kappa_a x^a$ to $f(z)$, and  without loss of generality claim that  the classical solution is
    \be
    \phi_{\rm cl}=f(z),
    \ee
where $f$ is a smooth function and
    \be
    z:= \sum_{a=0}^{d-1}\fK_a x^a,
    \ee
such that $\fK\cdot \fK=0$, i.e.,
    \be
    \fK_0^2=\sum_{a=1}^{d-1}\fK_a^2.
    \ee
 Thus the symmetry of the parameter space has two factors; $O(d-1)$ for the $\fK$-subspace and the diffeomorphism $z\to f(z)$.

\subsubsection{Classical solutions corresponding to  $\lambda_B>0$}
Now we can satisfy $\kappa\cdot\kappa=(4\pi G)^{-1}\lambda_B$ by choosing $\kappa=(0,\sqrt{\left|\kappa\cdot\kappa\right|},0,\cdots,0)$. In this way,  $\fK\cdot\kappa=0$ implies that $\fK_1=0$. Using this result in $\fK\cdot \fK=0$ we obtain $\fK_0=0$ for $d=2$ and
    \be
    \fK_0^2=\sum_{a=2}^{d-1}\fK_a^2,\hspace{1cm}{\rm for}\ \ \ d>2.
    \ee
Thus for $d=2$ the classical solution is $\phi_{\rm cl}=\kappa_0x^0+\kappa_1 x^1$ such that $\kappa_1^2-\kappa_0^2=(4\pi G)^{-1}\lambda_B$ and  similarly to $\lambda_B<0$, the parameter space  is symmetric under the action of the group $O(1,1)$.
For $d>2$ the classical solution is given by Eq.\eqref{phi-cl-general} and the symmetry of the parameter space has three factors; $O(d-1,1)$ for the $\kappa$-subspace, $O(d-2)$ for the $\fK$-subspace and  diffeomorphisms $z\to f(z)$.
\section{Concluding Remarks}\label{section-conclusion}
Second quantization of quantum field theory in four dimensional  Minkowski spacetime has been successful in modelling particle physics. Thinking about our universe as a four dimensional spacetime conceivably embedded in a higher dimensional geometry, it is reasonable to seek a quantum field theory describing particles localized to a nonstationary curved hypersurface embedded in a nonstationary curved spacetime.

In \cite{Loran:2018pxz} we have shown that a fermionic field theory exists in four dimensions with a consistent particle interpretation in general nonstationary curved spacetime  whose one-particle states are localized on two dimensional subspaces. In this work, we have shown that there exist scalar field theories whose one-particle states are well-defined in general $D$ dimensional nonstationary curved spacetimes  and their quanta are localized on $d\le D$ dimensional subspaces.  Therefore, in addition to providing a notion of scalar particles  in nonstationary curved spacetimes, this construction might be of some interest in the brane-world models  of a four dimensional universe embedded in a higher dimensional bulk, see, e.g., \cite{Csaki,Bajc:1999mh} and references therein. In particular, the ``massless'' quantum field theories localized on  two dimensional subspaces can be interpreted as $c=1$ conformal field theories embedded in the bulk.

The main obstacle to the particle interpretation of states of a quantum field theory in general  is the absence of a preferred notion of time translations in nonstationary spacetimes \cite{Wald-Book}. We have noted that  this problem can be circumvented by distinguishing the time-ordering of field operators arising in the path integral formalism from the timelike directions of the spacetime. The so-called  time-ordering occurring in the path integral can be attributed to the relative signs in the kinetic term. So, it can be separated from the spacetime geometry by using the action \eqref{action} whose field equation  does not engage the spacetime metric.

The path integral implies that the correlation function satisfies Eq.\eqref{DF-GCST-original}. We have been able to solve this equation, and explicate the path integral accordingly, by using a coordinate system denoted by $x$, in which the volume element $\fe\, d^Dy$ equals  $d^Dx$. Such coordinate systems are familiar in unimodular gravity \cite{Anderson,Padilla:2014yea}.  We observed that the Feynman propagator mimics the propagator of an auxiliary  scalar field confined to a $d$ dimensional flat hypersurface of a $D$ dimensional Minkowski spacetime. So, we  interpreted  it accordingly as the propagating amplitude of one-particle states localized on the $d$ dimensional hypersurface.

These scalars do not describe ordinary matter as can be seen from their stress tensor  $T_{\mu\nu}={\cal L}g_{\mu\nu}$
in which $\cL$ denotes the Lagrangian density.  They all act like perfect fluid with equation of state $w=-1$. By coupling to gravity, conservation of the stress tensor implies  that ${\cal L}$ is constant on-shell. So these scalars  add to the cosmological constant. We approached this problem classically and considered the on-shell value of $\cL$ as the bare cosmological constant $\lambda_B$.

The $d=1$ case is almost unique. Requiring the existence of nontrivial solutions to the classical field equation implies that the potential term is zero. So  the only freedom in writing the action is to choose a nowhere-vanishing vector field which is asymptotically timelike. In an FRW universe the vector field is timelike everywhere so the $d=1$ creeper  can be considered as a continuous  and monotonic time which does not commute with  $\lambda_B$ \cite{Smolin}. Furthermore, since in this model $\lambda_B<0$ it might be of some interest in the recent approaches to the cosmological constant problem \cite{Wang, Unruh}.

The existence of higher dimensional creepers depends on the number of linearly independent nowhere-vanishing vector fields in the spacetime which is a question in homotopy theory \cite{Adams}. Such creepers  are not unique because  we need to  choose the  potential term and also the (asymptotically) spacelike vector fields which together with the (asymptotically) timelike vector field  give the action. For example in an   FRW universe whose spatial sections are three dimensional spheres, $d>2$ dimensional creepers do not exist but we can  define $d=2$ dimensional creepers and for that purpose we need to choose one spacelike vector field from the three dimensional tangent space. Of~course,  it is reasonable to use the $SO(3)$ symmetry of the tangent space in order to consider all of these choices  equivalent to each other.

We studied $d\ge 2$ massless and  non-interacting  creepers and investigated  the parameter space of the classical solutions corresponding to a bare cosmological constant.    Contrary to the $d=1$ creepers, these models allow both positive and negative values for the bare cosmological constant with a signature of phase transition at $\lambda_B=0$. For $d>2$ the corresponding parameter spaces have different symmetries:  $O(d-1,1)$ for $\lambda_B<0$, $O(d-1)\times {\rm Diff}(\mathbb{R}^1)$ for $\lambda_B=0$ and $O(d-1,1)\times O(d-2)\times {\rm Diff}(\mathbb{R}^1)$ for $\lambda_B>0$. The symmetries of the parameter space for $d=2$ are $O(1,1)$ for $\lambda_B\neq 0$ and  $O(1)\times {\rm Diff}(\mathbb{R}^1)$ at $\lambda_B=0$.

\section*{Acknowledgments} The author is indebted to B.~Azad, S.~Lakzian, and M.~R.~Koushesh for fruitful discussions.

\appendix
\section{Foliation}\label{section-appendixA}
Here we  prove that the coordinates $y$ described in subsection \ref{section-revisit} exist. Let $\fI:=\{1,\cdots,\tilde{d}\}$ and $\fI^c=\{0,\cdots,D-1\}\setminus \fI$. Since the vector fields $\{v_i\}_{i\in\fI}$  commute with each other they are integrable and the corresponding field of $\tilde{d}$-planes are tangent  to the leaves of a  foliation $\{\sL_\alpha\}$ of the spacetime $\cM$. So every point  in  $\cM$  has a neighborhood $U$ and a system of local coordinates $\fz=(z^i,Z^I)$, $i\in \fI$ and $I\in\fI^c$ such that for each leaf $\sL_\alpha$, the components of $U\cap\sL_\alpha$ are described by the equation \cite{Foliat}
    \begin{align}
    &Z^I=\mbox{constant} &\mbox{for}\ \ \ I\in\fI^c.
    \end{align}
For $\sX(\sL_\alpha)$ denoting the set of vector fields on $\sL_\alpha$, and $v^{(\alpha)}_i:=\left.v_i\right\vert_{\sL_\alpha}$ we have
    \begin{align}
   & \sX(\sL_\alpha)={\rm span}\left\{v^{(\alpha)}_i\vert i\in\fI\right\}.
   \end{align}
Thus $ v^{(\alpha)}_i=\sum_{j\in\fI}{v_i^{(\alpha)}}^j\frac{\p}{\p z^j}$  where ${v_i^{(\alpha)}}^j:=\left.{\underline{v}_i}^j\right|_{\sL_\alpha}$, and ${\underline{v}_i}^j$ denote the components of the vector field $v_i$ in the $\fz$ coordinates, i.e. $v_i=\sum_{j\in\fI}{\underline{v}_i}^j\frac{\p}{\p z^j}$. Eq.\eqref{prop-1} and Eq.\eqref{prop-1-formula} imply that
    \be
    \label{A6}
    [v_i^{(\alpha)}, v_j^{(\alpha)}]=0.
    \ee
The cotangent vector field $\omega^i\in \Omega^1(\sL_\alpha)$ dual to  $v_i^{(\alpha)}$ is given by
    \begin{align}
    \label{omega}
       & \omega^i=\sum_{j\in\fI}{\omega_j}^i dz^j,
   \end{align}
 Since $\left\langle \frac{\p}{\p z^i},dz^j\right\rangle_{\sL_\alpha}:=\left\langle \frac{\p}{\p z^i},dz^j\right\rangle_{\cM}=\boldsymbol{\delta}^j_i$ we can  use the duality relation $\left\langle v_i^{(\alpha)},\omega^j\right\rangle_{\sL_\alpha}=\boldsymbol{\delta}^j_i$ to compute the components ${\omega_j}^i$ of $\omega^i$:
    \begin{align}
   \label{duality}
    &\sum_{k\in\fI}{v_i^{(\alpha)}}^k{\omega_k}^j =\boldsymbol{\delta}^j_i,&i,j\in\fI,
    \end{align}
which, in particular  implies that
    \begin{align}
    &\sum_{i\in\fI}{\omega_\ell}^i{v_i^{(\alpha)}}^k=\boldsymbol{\delta}_\ell^k,&k,\ell\in\fI.
    \end{align}
 This result together with Eq.\eqref{omega} gives
    \begin{align}
    &dz^i=\sum_{j\in\fI}{v_j^{(\alpha)}} ^i \omega^j,&i\in\fI.
    \label{differential}
    \end{align}
  Denoting the exterior derivative on $\sL_\alpha$ by ${\rm d}_\shortparallel:\Omega^1(\sL_\alpha)\to\Omega^2(\sL_\alpha)$ we have
    \bea
    {\rm d}_\shortparallel\omega^i&=&\sum_{m,n\in\fI}\left(\frac{\p}{\p z^m}{\omega_n}^i\right) dz^m \wedge dz^n\nn\\& =&\sum_{m,n,j,k\in\fI}\left(\frac{\p}{\p z^m}{\omega_n}^i\right){v_j^{(\alpha)}} ^m {v_k^{(\alpha)}} ^n \omega^j\wedge\omega^k\nn\\&=&-\sum_{m,n,j,k\in\fI}{\omega_n}^i{v_j^{(\alpha)}} ^m \left(\frac{\p}{\p z^m}{v_k^{(\alpha)}} ^n\right) \omega^j\wedge\omega^k,
    \eea
  where to obtain the second and third equalities we have used Eq.\eqref{differential} and Eq.\eqref{duality} respectively.
  Therefore
    \be
    {\rm d}_\shortparallel\omega^i=-\frac{1}{2}\sum_{j,k\in\fI}\left\langle \left[v_j^{(\alpha)},v_k^{(\alpha)}\right],\omega^i\right\rangle \omega^j\wedge\omega^k=0,
    \ee
  where we have used Eq.\eqref{A6}.
  Thus  $\omega^i$ is a closed form on $\sL_\alpha$. Assuming that the corresponding  cohomology group is trivial we conclude that $\omega^i$ is exact, i.e. there is a function $\sY^i=\sY^i(z,Z)$ such that $\omega^i=\left.{\rm d}_\shortparallel \sY^i\right\vert_{\sL_\alpha}$.   Eq.\eqref{omega} implies that
    \begin{align}
    \label{A1}
    &\frac{\p \sY^i}{\p z^j}={\omega_j}^i &\mbox{for}\ \ \ j\in\fI.
    \end{align}
  So there exist (nonunique) coordinate  transformation $z\to y=y(z)$ such that
      \begin{align}
      \label{A2}
      &y^i=\sY^i(z,Z)&\mbox{for}& \ \ \ i\in\fI,\\
      \label{A2-1}
      &y^I=y^I(Z)&\mbox{for}& \ \ \  I\in\fI^c.
      \end{align}
  The nonuniqueness encountered in Eq.\eqref{A2-1} is a reflection of  the concept of ``distinguished'' coordinates in \cite{Foliat}. The components ${{v}_i}^\mu$ of the vector field $v_i={v_i}^\mu\p_\mu$ in the $y$-coordinates are given by
    \be
    \label{A3}
    {{v}_i}^\mu=\sum_{k\in\fI}{\underline{v}_i}^k\frac{\p y^\mu}{\p z^k}.
    \ee
 Eq.\eqref{A2-1} implies that  ${{v}_i}^\mu=0$ for $\mu\in\fI^c$, and equations \eqref{A1}, \eqref{A2} and \eqref{duality} give
    \begin{align}
    \label{A4}
    &{{v}_i}^j=\sum_{k\in\fI}{\underline{v}_i}^k{\omega_k}^j=\boldsymbol{\delta}^j_i,
    \end{align}
    which results in Eq.\eqref{v-xf}.

Finally we study the one-dimensional foliation which is available in time-orientable spacetimes. In fact every open manifold (i.e. a manifold whose components are not  compact) and every compact manifold whose Euler characteristic is zero has a one-dimensional foliation  \cite{Foliat}. Suppose that $\fI=\{1\}$ and the vector field $v_1=\underline{v}(z,Z)\frac{\p}{\p z}$ is tangent  to leaves, where $z:=\fz^1$. We give the one-form dual to $v_1$ on $\sL_\alpha$ by  $\omega:=\underline{v}(z,Z)^{-1}dz$  noting that $v$ is a nowhere-zero vector field. Consequently
    \be
    \sY^1(z,Z)=\int^zd\zeta \underline{v}(\zeta,Z)^{-1}.
    \ee

\ed
Before studying the one-dimensional foliation and closing this section we give an alternate method to obtain the above result. Suppose that $\varpi^i\in T^*_\cP\cM$ is the one-form dual to $v_i$. In principle $\varpi^i=\varpi^i_\shortparallel+\varpi^i_\perp$ such that $\left<v_i,\varpi^i_\perp\right>_\cM=0$. So the duality relation
     \be
     \left\langle v_i,\varpi^j_\shortparallel\right\rangle_\cM=\left\langle v_i,\varpi^j\right\rangle_\cM=\boldsymbol{\delta}_i^j,
     \ee
 can be used to compute the components ${{\varpi_\shortparallel}_j}^i$ of $\varpi_\shortparallel^i=\sum_{j\in\fI}{{\varpi_\shortparallel}_j}^idz^j$ in terms of ${\underline{v}_i}^j$.
 together with Eq.\eqref{prop-1} would imply that
    \be
    {\rm d}_\shortparallel\varpi^i_\shortparallel:=\sum_{m,n\in\fI}\left(\frac{\p}{\p z^m}{{\varpi_\shortparallel}_n}^i\right) dz^m \wedge dz^n=0.
    \ee
Thus if $U$ is contractible to a point,  $\sY^i(z,Z)$ satisfying  $\varpi^i_\shortparallel={\rm d}_\shortparallel\sY^i(z,Z)$ exists and Eq.\eqref{v-xf} can be obtained.